\documentstyle[12pt,aasms4]{article}
\received{December 1995}
\accepted{13 march 1996}
\journalid{463}{1 Juni 1996}
\articleid{L1}{L4}

\begin{document}

\def\kms{km~s$^{-1}$}
\def\s{\ifmmode ^{\prime\prime} \else $^{\prime\prime}$ \fi}
\def\min{\ifmmode ^{\prime} \else $^{\prime}$ \fi}
\def\msun {M$_{\odot}$~}  \def\msund{M$_{\odot}$}  \def\mbh{$M_{\bullet}$}
\def\rbh{$r_{\bullet}$}
\def\es{elliptical galaxies{ }}
\def\etal{{\it et al.}{ }}
\def\mgs{Mg$_b-\sigma${ }}
\def\mg{Mg$_b$ line strength{ }}
\def\veldis{velocity dispersion{ }}
\def\lum{luminosity{ }}

\title{The Redshift Evolution of the Stellar Populations 
in Elliptical Galaxies\altaffilmark{1} }

\author{Ralf Bender\altaffilmark{2} and Bodo Ziegler\altaffilmark{2}}
\affil{Universit\"ats-Sternwarte, Scheinerstra\ss e 1,
        81679~M\"unchen, Germany  }

\author{Gustavo Bruzual}
\affil{ CIDA, Apartado Postal 264, M\'erida 5101-A, Venezuela }

\altaffiltext{1}{Partly based on observations carried out at the
European Southern Observatory, La Silla, Chile. }

\altaffiltext{2}{Visiting Astronomer of the German-Spanish Astronomical Center,
Calar Alto, operated by the Max-Planck-Institut f\"ur Astronomie,
Heidelberg, jointly with the Spanish National Commission for Astronomy}

\begin{abstract}

Velocity dispersions $\sigma$ and Mg absorption line-strengths Mg$_b$
have been measured for a sample of 16 ellipticals in 3 clusters at a
redshift of 0.37. Like local cluster ellipticals, these objects show a
correlation between Mg$_b$ and $\sigma$.  However, at any given
$\sigma$, the mean Mg$_b$ of the ellipticals at $z=0.37$ is
weaker than the mean Mg$_b$ of their local relatives in the Coma and Virgo
clusters. The Mg$_b$ weakening is smallest for the most luminous
ellipticals and larger for the fainter objects. This is unambiguous
evidence for {\it small but significant passive evolution} of the
stellar populations of elliptical galaxies with redshift.  It requires
that the bulk of the stars in cluster ellipticals has formed at
$z>2$. The most luminous objects may even have formed at $z>4$.
The Mg$_b-\sigma$ test is a very reliable estimator for the evolution of
old stellar populations because it is virtually independent from the
stellar initial mass function (IMF) and from the metallicities of the
galaxies. Furthermore, the influence of selection effects is minimal.

Consistent with the weakening of Mg$_b$ we find evidence for a B-band
luminosity evolution of about $0.5\pm 0.1$mag at $z=0.37$ from the
Faber-Jackson relation.  The combined information about the evolution
of Mg$_b$ and luminosity allows us to constrain both the slope of the IMF
in ellipticals and the cosmological deceleration parameter $q_o$. Our
present measurements are compatible with a standard Salpeter IMF and a
$q_o$ of $0.5\pm 0.5$.

\end{abstract}


\section{Introduction}

The redshift evolution of galaxies provides strong constraints on
their ages and formation processes as well as on theoretical models of
structure formation in the Universe. In the specific case of
elliptical galaxies, there is the additional prospect that, because of
their homogenous properties, it may eventually be possible to
calibrate their evolution accurately enough to allow their use as
cosmological standard candles or rods.

The luminosity and color evolution of elliptical galaxies or brightest
cluster members is generally found to be weak.  For various samples of
brightest cluster galaxies up to redshifts $z=1$, both optical and
infrared colors were found to change only very slowly with $z$
indicating a
high redshift of formation (e.g., Arag\'on-Salamanca \etal 1993,
Stanford \etal 1995). Similarly, Hamilton (1985) found only weak
evolution in the strength of the 4000\AA\ break. The bright end of the
galaxy luminosity function does not evolve significantly with redshift
either, being compatible even with a no-evolution scenario (e.g.,
Glazebrook \etal 1995, Lilly \etal 1995, Ellis \etal 1996).
Observations of distant galaxies up to $z=1$ from the ground and with
the Hubble Space Telescope revealed no significant deviations of the
surface brightness evolution from the Tolman relation, again an
indication of very small if any evolution (Sandage \& Perelmuter
1991, Dickinson 1995, Pahre et al. 1996). At lower redshifts ($z<0.4$),
Franx \& van Dokkum (1996) showed that the mass-to-light ratios of
ellipticals as obtained from the fundamental plane relation
(Djorgovski \& Davis 1987, Faber \etal 1987) change only very slowly
with redshift but in accordance with passive evolution of a very old
population.

These results are all indicative of a virtually passive evolution of most
cluster ellipticals up to $z\approx 1$. This may be in part a
selection effect because ellipticals that experience a merging or
accretion event may be transformed into an E$+$A-type (Dressler \&
Gunn 1983) or even
bluer object and therefore drop out of an ellipticals sample for some
time. But in general, these events cannot have injected a significant
fraction of young stars into brightest cluster ellipticals below
redshifts of 1. Otherwise, the observed redshift evolution would be
much stronger and, also, it would be difficult to explain the high
uniformity of colors and absorption line indices in {\it local}
cluster ellipticals (Bower \etal 1992, Bender, Faber \& Burstein
1993).  This does not contradict the claim that many ellipticals were
originally formed in {\it major} mergers (e.g., Schweizer 1990, Bender
1990b).  It simply means that, in rich clusters, the {\it major} merging and
star formation epoch is to be found at higher redshifts (e.g., Bender,
Burstein and Faber 1993, 1994). This picture has recently been shown
to be even in agreement with cold dark matter models (Kauffmann 1995).

With the redshift evolution of elliptical galaxies being very weak,
the challenge of measuring it accurately is correspondingly bigger.
In this paper we will analyse the evolution of ellipticals
with a method that is principally different from the methods
applied previously. Our test is based on the tight relation between
Mg-Index Mg$_b$ and velocity dispersion $\sigma$ of elliptical
galaxies and will be described in Section 2.  Section 3 briefly
describes sample selection and observations, Section 4 the data
analysis. Section 5 presents results and conclusions.

\section{Measuring the evolution of elliptical galaxies
with the Mg$_b-\sigma$ test}
 
A reliable measurement of the evolution of elliptical galaxies
requires a full understanding of selection effects. This means that
the mass and metallicity distributions of the distant galaxy sample
as well as of the local comparison sample have to be known very
accurately.  While the influence of the mass distribution is obvious,
the importance of metallicity results from the age--metallicity
degeneracy: {\it For stellar populations older than
$2\,$Gyr, an increase in metallicity by a factor of two and a decrease
in age by a factor of three results in almost identical optical and
near IR colors} (Worthey 1994).  Therefore, the colors of high
redshift ellipticals cannot be used to test their evolution as long as
their metallicities are not well constrained.

A method that automatically takes into account the effects of
different masses and metallicities is given by the Mg$_b-\sigma$ test
which we introduce and explain in the following.  Nearby cluster
ellipticals show a tight correlation between Mg$_b$ and $\sigma$
(e.g., Dressler et al. 1987). Because the Mg$_b$-index depends about
equally on metallicity and age, the tightness of the Mg$-\sigma$
relation implies that, at any given $\sigma$, the combined relative
spread in age and metallicity of cluster ellipticals must be smaller
than 15\% (Bender, Burstein \& Faber 1993).  Therefore, measuring the velocity
dispersion of a local elliptical allows to constrain its metallicity
to better than 15\%. Moreover, if ellipticals evolve essentially
passively below $z\approx 1$, then the same applies to distant
ellipticals. {\it I.e.}, we can assume that all ellipticals up to
$z\approx 1$ which have similiar velocity dispersion also have
similiar metallicities.  Any reduction of Mg$_b$ of distant
ellipticals must therefore be due to lower age.  Consequently, the
observation of the Mg$-\sigma$ relation as a function of redshift
gives a straightforward estimate of the evolution of the stellar
populations in elliptical galaxies.

The age determination with the Mg$-\sigma$ test has the additional
advantage that, unlike the evolution of luminosity or surface
brightness, {\it the evolution of Mg$_b$ line strength is virtually
independent from the stellar initial mass function} (e.g., Worthey
1994). Finally, we note that small differences in the Mg$-\sigma$
relation which might exist between low and high density environments
(e.g., de Carvalho \& Djorgovski 1992, Lucey 1995) do not invalidate
our conclusions in the following. This is because the elliptical
galaxies we compare at different redshifts all live in the centers of
medium rich to rich clusters.

\section{Sample selection and observations}

The distant elliptical galaxies selected for observation were members
of three clusters of galaxies at similar $z$: Abell~
370 ($z=0.375$), MS~1512$+$36 ($z=0.372$), and CL~0949$+$44 ($z=0.377$). 
The redshift of 0.37 was chosen
because it allows a good determination of Mg absorption strengths
undisturbed by strong telluric emission and absorption lines. The
color selection criterion for the galaxies was $\rm (B-V)_{\rm rest
frame} >0.8$, in turn being equivalent to $\rm Mg_b > 3$\AA\ (using
relations between $\rm B-V$, Mg$_2$ and Mg$_b$ from Burstein \etal
1984 and Bender \etal 1993). Colors and magnitudes
were taken from Pickles \& van der Kruit (1991) or derived from our own CCD
photometry carried out with the Calar Alto 2.2m telescope.  Generally, in
each cluster the two to four brightest galaxies plus a random sample
of fainter objects (obeying however the color constraint) were selected
for spectroscopic observation.
 
The spectroscopy was carried out at the 3.5m telescope on Calar Alto,
Spain, and at the ESO NTT on La Silla, Chile, during in total 26
nights between December 1993 and July 1995.  We used the Boller
\& Chivens double spectrograph at the Calar Alto and the EMMI
focal reducer at ESO.  The
redshifted galaxies and the local template stars were observed in the
same rest wavelength range between about 4500\AA\ to 6000\AA\ and with
identical spectral resolution ($\sigma_{\rm instr} \approx
100\,$km/s). The aperture size was 3.5'' by 3.5''. Exposure times
between 8 and 12 hours were needed to reach a signal-to-noise ratio of
at least 40 per \AA\ in the spectra of the galaxies.

\section{Data analysis}

{\bf Spectroscopy.}
The CCD spectra were reduced in the standard way (e.g., Bender, Saglia
\& Gerhard 1994). Special care was taken to achieve accurate sky
subtraction. The flux standards were used to control the strength of
weak telluric H$_2$O absorption bands around 7200{\AA}.  The velocity
dispersions of the galaxies were derived using the Fourier Correlation
Quotient method (Bender 1990a). The Mg$_b$ equivalent width was
determined as described in Faber \etal (1985). Mg$_b$ was corrected
for differences in spectral resolution, for velocity dispersion
broadening and for redshift dependent factors.  Because the spectra of
the $z=0.37$ ellipticals were integrated over a much larger physical
surface area than the spectra of the nearby comparison ellipticals in 
the Coma and Virgo cluster (taken from Dressler \etal 1987), an aperture
correction had to be applied. We zeroed the correction to a 4'' by 4''
projected aperture size at the distance of the Coma Cluster. The
aperture correction was computed from velocity dispersion profiles and
Mg$_b$ profiles published in, e.g., Davies \etal (1993), Gonzalez
(1993), Carollo \etal (1995) and Fisher \etal (1995).

Finally, the Mg$_2$ absorption magnitudes of the local comparison
sample of cluster ellipticals in Coma and Virgo (Dressler \etal 1987)
had to be transformed to Mg$_b$ equivalent widths because for the
distant ellipticals the Mg$_b$ values could be determined with much
better S/N than Mg$_2$ values. For that purpose we used the very good
Mg$_b$-Mg$_2$ correlation, empirically given in Burstein \etal (1984)
and Gonzalez (1993).

\medskip
{\bf Imaging.} In order to analyse the Faber-Jackson relation for
the redshifted ellipticals, their rest-frame total luminosities in the
B-band had to be calculated. For this purpose, we first extrapolated
the observed BVRI {\it aperture} magnitudes to {\it total} magnitudes
on the basis of the ratio between the galaxies' effective radii $r_e$
and the aperture size. An estimate of the $r_e$ was obtained using the
$r_e-\sigma$ relation of nearby ellipticals. Typical aperture
corrections ranged between 0.2 and 1.1$\,$mag, with an error of
about 0.2$\,$mag (as estimated from the scatter of the local $r_e-\sigma$
relation).  In the second step, rest-frame $B$ luminosities were
calculated from fits of spectral energy distributions (from Bruzual \&
Charlot 1996) to the observed BVRI values.

\section{Results and Discussion}

{\it Figure 1} shows the \mgs relation for cluster ellipticals at a
redshift of $z = 0.37$ in comparison to local elliptical galaxies in
the Coma and Virgo cluster. Two main conclusions can be drawn:

\placefigure{fig1}

 (a) The distant ellipticals also show a correlation between Mg$_b$
index and velocity dispersion $\sigma$ as local ellipticals do. 
The slope of the \mgs relation at higher redshift appears
to be slightly steeper than today indicating that low luminosity
ellipticals may be systematically younger than high luminosity
ellipticals. This may support a recent claim by Faber \etal (1995)
according to which the mean ages of ellipticals should systematically
decrease with decreasing luminosity. The distant objects with
{\it very} low Mg$_b$ could be genuinely younger or may only have
experienced a more recent but smaller star formation event.

 (b) There is clear evidence for evolution: at any given velocity
dispersion, the mean Mg$_b$ of ellipticals at $z=0.37$ is
significantly weaker than at $z=0$. The effect is relatively weak for
big ellipticals and larger for small ellipticals.  On average,
the Mg$_b$ absorption is reduced by about 0.3~\AA. Note that this
conclusion is not affected by selection biases since our
color selection criterion (see Section 3) only cuts off ellipticals
with Mg$_b < 3\,$\AA.  

\smallskip
On the basis of Worthey's (1994) population synthesis models we can
use this weaking of Mg$_b$ to estimate the ratio of the mean age of
the $z=0.37$ ellipticals relative to the mean age of local
ellipticals. We derive the following dependence of Mg$_b$ on age $t$
and metallicity $Z$: $\log$~Mg$_b = 0.37 + 0.20~\log t + 0.31~\log
Z/Z_o$ (valid for $t > 5\,$Gyrs and $1/3 < Z/Z_\odot < 3$). The
proportionality factors of $\log t$ and $\log Z$ are very robust and
only weakly dependent on IMF and on uncertainties in the population
synthesis models (Bruzual 1996).  The small evolution we
see in {\it Figure 1} strongly supports the notion that ellipticals have
evolved mostly passively between $z=0.37$ and $z=0$. Since then
$Z \approx const.$, the observed reduction
of Mg$_b$ can be directly translated into a relative age difference.
The three dashed lines in {\it Figure 1} give the {\it expected} \mgs
relation at $z=0.37$, if the mean formation redshifts of the stars in
ellipticals were formed at $z_f = 4.5, 2, 1$, respectively. Independent from
H$_o$ and only weakly dependent on q$_o$ we conclude that {\bf the
bulk of the stars in the {\it luminous} cluster ellipticals must have
formed at redshifts} $\bf z>2$, the higher luminosity objects may even
have formed at $\bf z>4$. Because of the possible presence of some
younger stars injected in minor accretion events and because we
observe luminosity-weighted spectra, these age estimate is rather a
lower limit.

\placefigure{fig2}

\smallskip
The age-driven reduction of the Mg$_b$ absorption in ellipticals at higher
redshifts should correspond to an increase in their
luminosity. This should be detectable via the Faber-Jackson relation
(Faber \& Jackson 1976).  In order to derive the Faber-Jackson
relation for our distant cluster ellipticals, their rest-frame $B$
luminosities were calculated as described in Section 4 and using $H_o=
50$~km/s/Mpc and $q_o=0.5$.

The Faber-Jackson relation for the $z=0.37$ cluster ellipticals and
for the Coma and Virgo ellipticals is shown in the upper panel of {\it
Figure 2}. It is evident that the distant ellipticals are brighter
than their local counter-parts by about 0.5$\pm 0.1\,$mag in the
B-band.  Is this consistent with the weakening of Mg$_b$ in the
Mg$-\sigma$ relation? Using Worthey's models with a Salpeter IMF, we
obtain that age variations lead to a correlated change of the B-band
luminosity with the Mg$_b$ value according to $\rm \Delta  =
1.2 \Delta\, Mg_b/$\AA. Thus, the evolution of the B \lum as estimated
from the weakening of Mg$_b$ should be $\Delta B$($z\approx0.4$)
$\approx -0.4$ mag, consistent with the actually observed brightening.
This corresponds to an evolution in rest frame $(B-V)$ of $-0.04\,$
mag and $(V-K)$ of $-0.12\,$mag, in accordance with the measurements
of Stanford et al. (1995).

We can also turn the above procedure around and correct the luminosity
of each galaxy by its individually calculated evolution correction
$\rm \Delta M_B = 1.2 \Delta\, Mg_b/$\AA. The lower panel of
{\it Figure 2} shows the result: the $z=0.37$ ellipticals and the
Coma and Virgo ellipticals now all fall on top of each other. 
Even the slope of the Faber-Jackson relation at
$z=0.37$ is similar within the errors to the locally observed slope.

We conclude that the evolution of the stellar populations in
ellipticals as derived from the \mgs relation and from the
Faber-Jackson relation are consistent. Within the limits of our errors,
there is no evidence for a very unusual slope of the IMF, nor a very
unusual value of the cosmological constant. Changing the slope of the
IMF by $\Delta x =1$ would cause a change of the luminosities at
$z=0.37$ by ca. $\pm 0.13$mag (Bruzual \& Charlot 1996 models). Similarly,
changing $q_o=0.5$ by $\pm 0.5$ causes a change of the luminosities by
ca. $\pm 0.11$mag.

In a future paper, we will combine our spectroscopic data with HST
photometry. This will allow us to investigate the evolution of the
elliptical galaxies more accurately by employing the fundamental plane
relations. In combination with the Tolman surface brightness test, one
may then also be able to obtain better and, more importantly,
independent constraints on the slope of the IMF and on the value of
$q_o$.

\acknowledgements

We thank the Calar Alto and La Silla staff for efficient support
during the observations. Discussions with Drs. P. Belloni,
L. Greggio, U. Hopp, G. Kauffmann, A. Renzini and R. Saglia are
gratefully acknowledged.  Thanks also go to Dr. R. Carlberg who
provided us with a list of red members of MS~1512$+$36.  The work of
R.B. and B.Z. was supported by SFB 375 of the DFG and by the MPG.

\clearpage

\figcaption[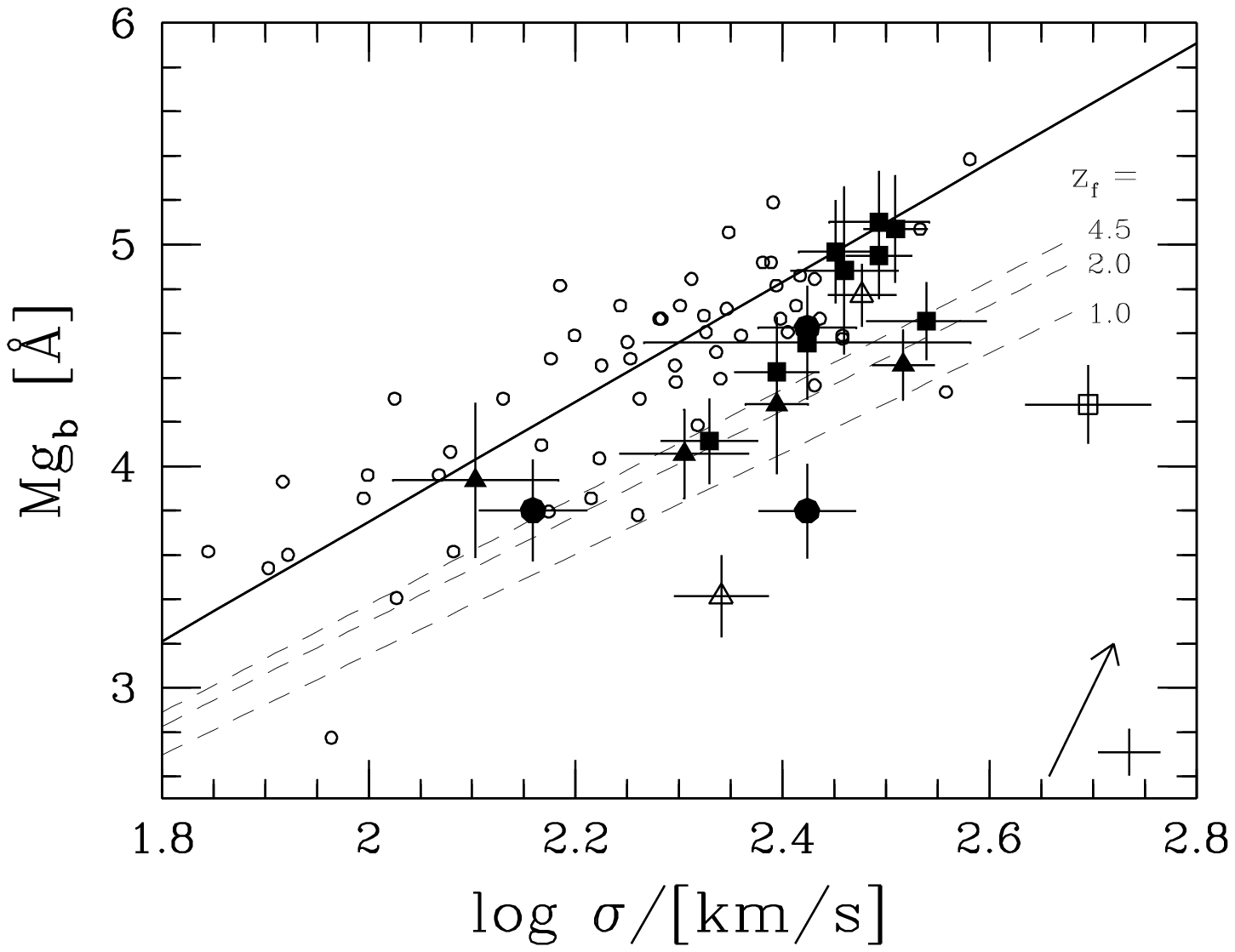]{Mg$_b-\sigma$ relation for
ellipticals in the Coma and Virgo cluster (small open circles,
typical error bar in the lower right corner) and for ellipticals at
$z=0.37$ in Abell~370 (squares), MS~1512$+$36 (triangles) and CL~0949$+$44
(circles). For the latter,
filled symbols refer to reliable measurements within the
errorbars, open symbols to systematic uncertainties larger than the error bars
and due to emission lines and/or night sky subtraction
problems. The solid line is a fit to the Virgo and Coma ellipticals,
the dashed lines correspond to the expected \mgs relation at $z=0.37$
for a mean formation redshift of the stars in ellipticals at $z_f =
4.5, 2, 1$ (top to bottom; $q_o=0.5$). The arrow in the lower right
corner gives the size of the aperture correction that has been applied
to the distant ellipticals. \label{fig1} }

\figcaption[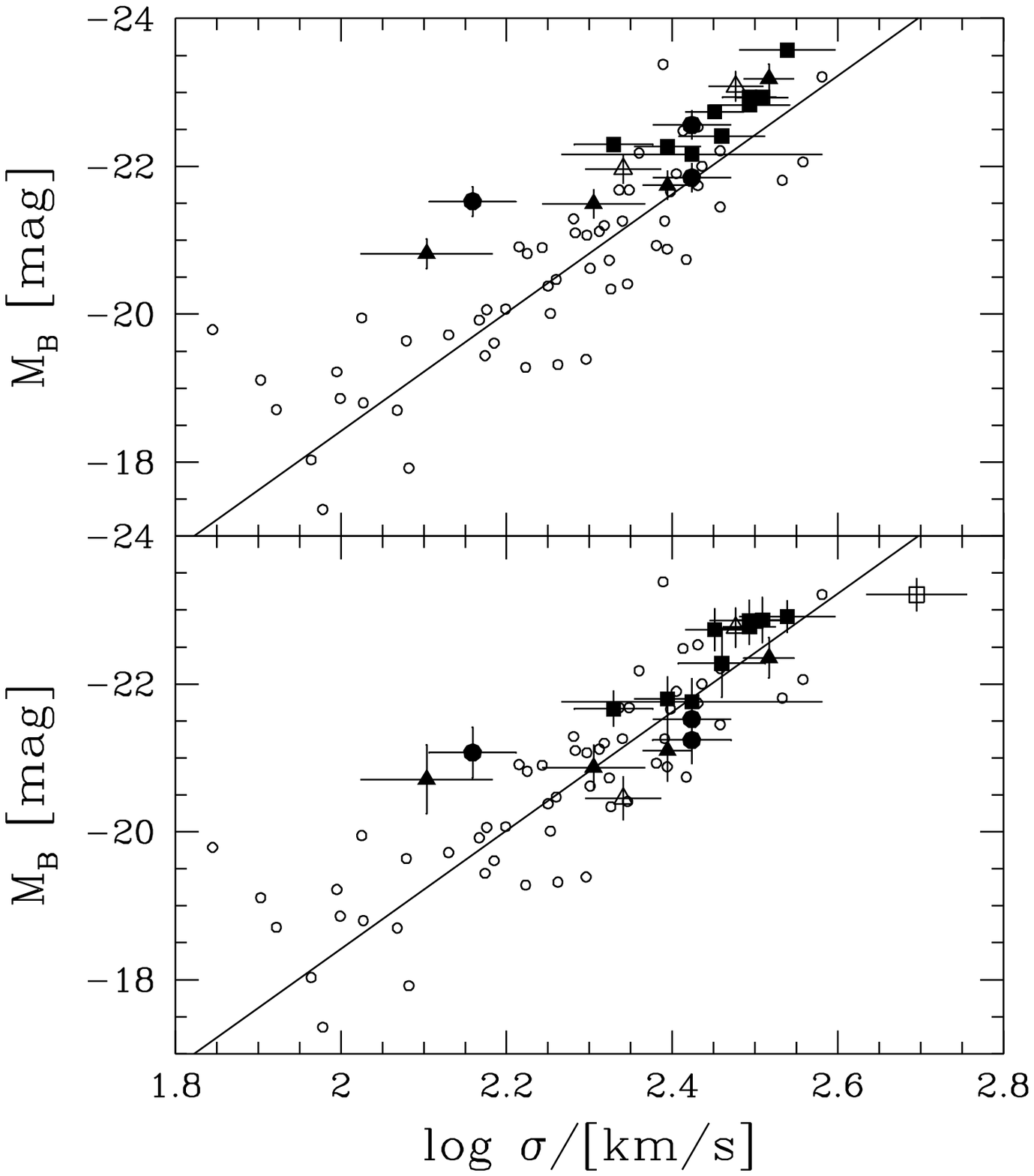]{The Faber-Jackson relation
for Coma and Virgo ellipticals (small dots) and for
ellipticals  at $z=0.37$ ($q_o = 0.5$, symbols as in F{\sc IG}.~1); 
upper panel: rest-frame M$_B$, lower panel: M$_B$ of
$z=0.37$ ellipticals corrected for evolution (see text for
explanation). The solid line is a fit to the Virgo and Coma
ellipticals. \label{fig2} }

\clearpage

\begin{figure}
\figurenum{Figure 1}
\plotone{fig1_final.eps}
\end{figure}

\clearpage

\begin{figure}
\figurenum{Figure 2}
\plotone{fig2_final.eps}
\end{figure}

\end{document}